\documentstyle[aps,preprint,prb]{revtex}
\begin{document}
\title{\bf Weak Localization Effect in Superconductors}
\author{Yong-Jihn Kim and K. J. Chang}
\address{Department of Physics,  Korea Advanced Institute of Science and 
Technology,\\
 Taejon 305-701, Korea}
\maketitle
\begin{abstract}
We study the effect of weak localization on the transition temperatures
of superconductors using time-reversed scattered state pairs, and find
that the weak localization effect weakens electron-phonon interactions.
With solving the BCS $T_{c}$ equation, the calculated values for $T_c$
are in good agreement with experimental data for various two- and
three-dimensional disordered superconductors.
We also find that the critical sheet resistance for the suppression
of superconductivity in thin films does not satisfy the universal behavior
but depends on sample, in good agreement with experiments.

\end{abstract}

\vfill\eject
\section{Introduction}
Since the scaling theory of Anderson localization,$^{1}$ our understanding
of the electronic properties of disordered conductors has been 
advanced considerably.$^{2,3}$
However, the effect of localization on superconductivity is still not well
understood.$^{4-7}$ A unified picture for the disorder
effects on the superconducting temperature ($T_{c}$) and the critical
field ($H_{c2}$) is still lacking.
In the presence of disorder, it was even claimed that the Eliashberg
theory breaks down for two-dimensional superconductors.$^{8}$
There have been many experimental studies to explain a competition
between localization and superconductivity in two- and three-dimensional 
systems.$^{8-16}$
In homogeneous amorphous thin films, an empirical formula$^{5}$ showed
that the reduction of $T_{c}$ is proportional to the sheet resistance
$R_{\Box}$.
For bulk amorphous $\rm{InO_{x}}$, Fiory and Hebard$^{13}$ found that
both the normal-state conductivity $\sigma$ and the critical temperature 
vary as $(k_{F}\ell)^{-2}$ due to the localization effect,
where $k_{F}$ and $\ell$ are the Fermi wave vector and the elastic
mean free path, respectively.
In A15 superconductors such as $\rm{Nb_{3}Sn, V_{3}Si,}$ and
$\rm{V_{3}Ga}$, the universal degradation of $T_{c}$ with impurities
was also found,$^{17}$ following the variation of $\rho^{2}\propto
(k_{F}\ell)^{-2}$,$^{18-20}$ where $\rho$ is the normal-state resistivity.
Previously, the decrease of $T_c$ with increasing of disorder was attributed
to the enhanced Coulomb repulsion $\mu^{*}$.$^{4,17,21,22}$
However, tunneling experiments$^{8, 23-25}$ did not support such an
argument and indicated instead a decrease of the electron-phonon
coupling $\lambda$.$^{8,23-26}$
As an alternative explanation, a singularity of the density of states
at the Fermi level caused by long-range Coulomb interactions was suggested
to change $T_{c}$'s for Pb and Sn thin films.$^{27}$
However, it was pointed out that the singularity of the density of states
does not affect thermodynamic quantities such as $T_{c}$.$^{28}$
In addition, it was shown experimentally that the effect of the Coulomb gap
is minor in superconductors.$^{29,30}$

In this paper we present the results of theoretical studies on the
weak localization effect on the superconducting temperatures of
disordered superconductors.
Using time-reversed scattered state pairs, we are able to explain
available experimental data, based on a unified picture that
the electron-phonon coupling decreases with increasing of disorder.
Our theory also explains other experimental results that in disordered
superconductors $T_{c}$ can be enhanced by spin-orbit
scatterings,$^{12,31-33}$ which results from the anti-localization effect.
In thin films, we suggest that the critical sheet resistance for the
suppression of superconductivity is not a universal constant but
a sample-dependent quantity.
Some preliminary results were reported elsewhere.$^{34}$

\section{Theory}

It has been recently realized that there exist localization corrections
in electron-phonon interactions.
To find these correction terms,$^{35-38}$ it is a prerequisite to
understand the limitation of Anderson's theorem and the pairing problem 
in Gor'kov's formalism and Bogoliubov-de Gennes equations.
The Anderson's theory in dirty superconductors$^{39}$ is based on
the fact that the exact eigenstates in the presence of impurities
consist of time-reversed degenerate pairs;
the scattered state $\psi_{n}$ of an electron with spin up is paired
with the other electron with spin down in the time-reversed state
$\psi_{\bar n}$.
Then, the reduced Hamiltonian in scattered-state representation is
written as
\begin{eqnarray}
H^{'}_{red}=\sum_{nn'}V_{nn'}c_{n'}^{\dagger}c_{\bar n'}^{\dagger}
c_{\bar n}c_{n},
\end{eqnarray}
where $c_{n}^{\dagger}$ and $c_{n}$ are creation and annihilation
operators, respectively, for an electron in the state $\psi_{n}$.
Assuming a point coupling $-V\delta({\bf r}_{1}-{\bf r}_{2})$ for
phonon-mediated electron-electron interactions, the matrix elements
$V_{nn'}$ are expressed as$^{40}$
\begin{eqnarray}
V_{nn'}=-V\int\psi_{n'}^{*}({\bf r})\psi_{\bar n'}^{*}({\bf r})
\psi_{\bar n}({\bf r}) \psi_{ n}({\bf r})d{\bf r}.
\end{eqnarray}
If the scattered state $\psi_{n\sigma}$ is expanded in terms of plane
waves $\phi_{{\vec k}\sigma}$, 
\begin{eqnarray}
\psi_{n\sigma}=\sum_{\vec k}\phi_{{\vec k}\sigma}\langle {\vec k}|n\rangle,
\end{eqnarray}
$V_{nn'}$ is rewritten as$^{41}$
\begin{eqnarray}
V_{nn'}=-V(1+\sum_{{\vec k}\not= -{\vec k'},{\vec k'},{\vec q}}\langle -{\vec k'}|
n\rangle \langle{\vec k}|n\rangle^{*} \langle {\vec k}-{\vec q}|n'\rangle
\langle {\vec k}-{\vec q}|n'\rangle
\langle -{\vec k'}-{\vec q}|n'\rangle^{*}).
\end{eqnarray}
The second term in Eq. (4) is negligible in dirty limit, where the mean
free path $\ell$ is in the range of 100 \AA, while its contribution is
meaningful in weak localization limit, where $\ell$ is of the order of 10 \AA.
Consequently, the Anderson's theorem is only valid in the dirty limit.

In order to calculate the correction term in Eq. (4), we need information
on the weakly localized scattered states.
Kaveh and Mott$^{42}$ derived these scattered states in the form of
power-law and extended wavefunctions for both two and three dimensional
systems.
Haydock$^{43}$ also showed that the asymptotic form of the scattered states
is power-law like in weakly disordered two-dimensional systems.
Here we use the scattered states of Kaveh and Mott$^{42}$ to calculate
the matrix elements $V_{nn'}$.
In this case, since we are dealing with the bound state of a Cooper pair
in a BCS condensate, only the power-law wavefunctions within the BCS coherence
length $\xi_{o}$ are relevant.$^{44}$
In weak localization limit, the effective coherence length is reduced to
$\xi_{eff}\approx \sqrt{\ell\xi_{o}}$.
It is important to take into account the size of the Cooper pairs 
for self-consistent calculations of the matrix elements.
A similar situation also occurs for the localized states, which are not
very sensitive to the change of the boundary conditions.$^{45}$
Because of the power-law-like wavefunctions, the correction term
due to weak localization represents physically the decrease of
the amplitudes of the plane waves.
Thus, the Cooper-pair wavefunctions basically consist of the plane 
waves with reduced amplitudes.
Including the weak localization effect, the resulting matrix elements $V_{nn'}$
for two- and three-dimensional systems are written as
\begin{eqnarray}
V_{nn'}^{2d}\cong 
-V[1-{2\over \pi k_{F}\ell}ln(L/\ell)],
\end{eqnarray}
\begin{eqnarray}
V_{nn'}^{3d}\cong
-V[1-{3\over(k_{F}\ell)^{2}}(1-{\ell\over L})].
\end{eqnarray}
Because of the impurity effect, the electron-phonon interactions are
weakened, as clearly seen in Eqs. (5) and (6).

\section{Results and discussion}

If we assume that the electron-phonon coupling constant is modified
by the weak localization effect, the effective coupling constant
$\lambda_{eff}$ from Eq. (5) for three-dimensional systems can be written as
\begin{eqnarray}
\lambda_{eff}=\lambda[1-{3\over(k_{F}\ell)^{2}}(1-{\ell\over L})].
\end{eqnarray}
Then, using the modified BCS $T_{c}$ equation,
\begin{eqnarray}
k_{B}T_{c} = 1.13 \hbar \omega_{D} exp(-1/\lambda_{eff}),
\end{eqnarray}
the change of $T_{c}$ relative to $T_{co}$ (for a pure metal) can be easily
estimated to first order in the weak localization correction term of the 
coupling constant;
\begin{eqnarray}
{T_{co}-T_{c}\over T_{co}} \cong {1\over \lambda}{3\over (k_{F}\ell)^{2}}(
1-{\ell\over L}).
\end{eqnarray}

For a metal with $\omega_{D}$ = 300 K, $k_{F}$ = 1 \AA$^{-1}$,
and $L$ = 1000 \AA$/T$, the variations of $T_c$ are plotted as a function
of $1/k_{F}\ell$ in Fig. 1, assuming $T_{co}$ = 4 and 12 K.
The inelastic mean free path results from electron-electron scatterings,
and the correction term in Eq. (9) is negligible for $L \gg \ell$. 
We find that $T_{c}$ changes slowly with increasing of $1/k_{F}\ell$
until $1/k_{F}\ell$ equals to 0.1, in good agreement with experimental
results.$^{13,18-20}$
This behavior which satisfies the Anderson's theorem is attributed to
the fact that the change of $T_c$ is proportional to $\rho^{2}$, i.e.,
the square of impurity concentration. 
However, for $1/k_{F}\ell>0.1$, $T_c$ decreases more significantly 
due to the weak localization effect caused by ordinary impurities.
In this weak localization limit, the Anderson's theorem is not valid. 
Our theoretical results are also compared with experimental data$^{46}$
for A15 superconducting materials in Fig. 2.
Using the same values of $k_{F}$ = 0.87 \AA$^{-1}$ and $L$ = 1000 \AA$/T$
for Nb$_3$Ge and V$_3$Si with $\omega_{D}$ = 302 and 330 K and
$T_{co}$ = 23 and 17 K, respectively, we find good agreements between
theory and experiment.
In this case, since it is difficult to evaluate $k_{F}\ell$
up to a factor of 2,$^{47}$ we assume that $\rho = 100 \mu \Omega cm$
corresponds to $k_{F}\ell$ = 3.3 and 3.75 for Nb$_3$Ge and V$_3$Si,
respectively.

Similarly, we can write down the effective electron-phonon coupling constant
for two-dimensional systems such as
\begin{eqnarray}
\lambda_{eff}=\lambda[1-{2\over\pi k_{F}\ell}ln(L/\ell)].
\end{eqnarray}
Since $1/k_{F}\ell$ is related to $R_{\Box}$ by the Drude formula,
the change of $T_c$ is expressed as,
\begin{eqnarray}
{T_{co}-T_{c}\over T_{co}} \cong {1\over \lambda}
{e^{2}\over \pi^{2}\hbar}R_{\Box}ln(L/\ell),
\end{eqnarray}
and this formula well satisfies the empirical relationship between
$T_{c}$ and $R_{\Box}$ for two-dimensional superconductors. 
In Fig. 3, the variations of $T_{c}$ with $1/k_{F}\ell$ are drawn
for two superconductors with $T_{co}$ = 4 and 8 K, assuming the same
values of $\omega_{D}$ = 300 K, $\ell$ = 4 \AA, and $L$ = 1000 \AA $\sqrt{T}$.
The inelastic mean free path obtained from disordered two-dimensional
systems$^{48}$ are employed, with the $1/\sqrt{ln T}$ dependence removed.
In contrast to the 3-dimensional case, $T_{c}$ varies linearly with
increasing of impurity concentration, and the initial slope of $T_{c}$
depends on superconductor because of the prefactor $1/\lambda$ in Eq. (12).
Thus, the critical sheet resistances for the suppression of superconductivity
in thin films do not provide
the universal behavior, in good agreement with experiments.$^{5,9,12}$

Since the inelastic mean free path depends on temperature,
the effective electron-phonon coupling also varies with
temperature.
Including the temperature effect on $\lambda_{eff}$, the variation
of the gap parameter with temperature is plotted for two different
values of $1/k_{F}\ell$ = 0.039 and 0.078 in Fig. 4.
The gap parameters are found to be lower than those obtained using
the $T$-independent $\lambda_{eff}$'s.
Experimentally, this behavior may be observable for lower $T_{c}$
superconductors, while strongly-coupled superconductors such as Pb and
Nb may not be appropriate because the BCS model is not applicable
for these systems.

For Mo-C$^{49}$ and a-MoGe$^{12}$ thin films, our calculated $T_{c}$'s
are plotted as a function of $1/k_{F}\ell$ and compared with experimental
data in Fig. 5.
Here we employ the Drude formula to represent the decrease of $\ell$
when the sheet resistance $R_{\Box}$ increases.
As in the 3-dimensional case, because of the difficulty in evaluating
$1/k_{F}\ell$ up to a factor of 2 from experimental data, we assume $\ell$ = 
3.5 \AA \ and\ $L$ = 2000 \AA$\sqrt{T}$ for best fit in the Mo-C sample.
For the a-MoGe film, the use of $\ell$ = 4.0 \AA and $L$ = 1000
\AA$\sqrt{T}$ is found to give the best agreement with experiment.
In this case, the Drude formula is slightly adjusted by the relation
$1/k_{F}\ell$ = 1.6667 $(e^{2}/2\pi\hbar) R_{\Box}$.
Similarly to the three-dimensional case, we find that the critical sheet
resistance for the suppression of superconductivity does not follow the 
universal behavior.

We can also examine the effect of weak localization on superconductivity,
using the strong or weak coupling Green's function theory.
In previous approaches,$^{50-52}$ because the Anderson's time-reversed
scattered-state pairs were not employed, the correction term due to
weak localization was missing in the electron-phonon interaction.
The Green's function formalism leads to the pairing states 
formed by a linear combination of the scattered states.$^{35-38}$ 
Since these are still extended states, we do not expect the weak
localization effect.
Using the Anderson's pairing condition for the strong coupling equation
and the Einstein model for the phonon spectrum, the gap equation can be
written as,$^{37}$ 
\begin{eqnarray}
\Delta(n,\omega)=\sum_{\omega'}\lambda(\omega-\omega')\sum_{n'}V_{nn'}
{\Delta(n',\omega')\over \omega'^{2}+\epsilon_{n'}^{2}},
\end{eqnarray}
where
\begin{eqnarray}
V_{nn'}=-V\int|\psi_{n}({\bf r})|^{2}|\psi_{n'}({\bf r})|^{2}d{\bf r},
\end{eqnarray}
\begin{eqnarray}
\lambda(\omega-\omega')={\omega_{D}^{2}\over \omega_{D}^{2}+
(\omega-\omega')^{2}}.
\end{eqnarray}
In this case, the use of $V_{nn'}$ in Eq. (2) can also give rise
to a modification of the electron-phonon interaction due to impurities
in the strong coupling theory. 
In previous theories,$^{50-52}$ however, the electron-phonon interaction
remains unchanged, even if the wavefunctions are localized.

It is interesting to see the superconductor-to-insulator transitions
in ultrathin films, which usually occur by changing film thickness or applying
magnetic fields.$^{49,53,54}$ 
Previously, a dirty boson theory$^{55}$ was used to explain
this experimental feature, implying that the critical sheet resistance
for the suppression of supercondcutivity is a universal constant,
$h/4e^{2}$ = 6.45 K$\Omega$.
However, although it is difficult to determine experimentally a well-defined
value for the critical sheet resistance $R^{c}_{\Box}$ because the
transitions do not occur sharply, the measured values for $R^{c}_{\Box}$
were shown to depend on sample.$^{49,53,54}$
We point out that the weak localization effect considered here
is still a dominant contribution to the suppression of superconductivity
over other higher order terms beyond the weak localization regime.$^{2,3}$
In fact, Fig. 3 shows that $T_{c}$ is completely suppressed in the region
of $k_{F}\ell \approx 6\ {\rm or}\ 7$, which is still in the weak
localization regime. 
Thus, our results indicate that the superconductor-insulator transition
is not a sharp phase transition but a crossover phenomena from quasi-two
dimensional to two-dimensional.

Finally, we suggest that if the decrease of $T_c$ is caused by weak
localization in disordered superconductors, adding impurities with
large spin-orbit couplings will compensate for this decrease.
In fact, such behavior was observed for several 3-dimensional
samples,$^{12,31-33}$ while it needs to be tested for 2-dimensional 
case.  
We expect that the critical sheet resistance is increased
by enhancing the spin-orbit scattering. 
It is known that magnetic fields suppresses the weak localization effect.
In this case, however, since the magnetic field decreases $T_{c}$,
the $T_{c}$ decrease in a pure sample should be compared with that
of a weakly disordered one.
If the difference would exist, it is suggested to result from the weak 
localization effect, then, the critical sheet resistance will also change
with increasing of magnetic field.

\section{Conclusions}

In conclusion, we have studied the effect of weak localization on
superconductors within the BCS theory.
We find that the weak localization decreases the electron-phonon coupling
constant, thereby, suppressing $T_{c}$. 
The calculated variations of $T_{c}$ with increasing of impurity
concentration are found to be in good agreement with experiments
for both 2- and 3-dimensional systems.
The recovery of $T_{c}$ with impurities having large spin-orbit scatterings
supports strongly our theory.
We suggest that the critical sheet resistance for the suppression 
of superconductivity in thin films is not a universal constant, but
a sample-dependent quantity, in good agreement with experiments.

\vspace{0.5cm}
\acknowledgments
We are grateful to Yunkyu Bang, Hu Jong Lee, and H. K. Sin for
helpful discussions.
This work is supported by the brain pool project of KOSEF and
the MOST.

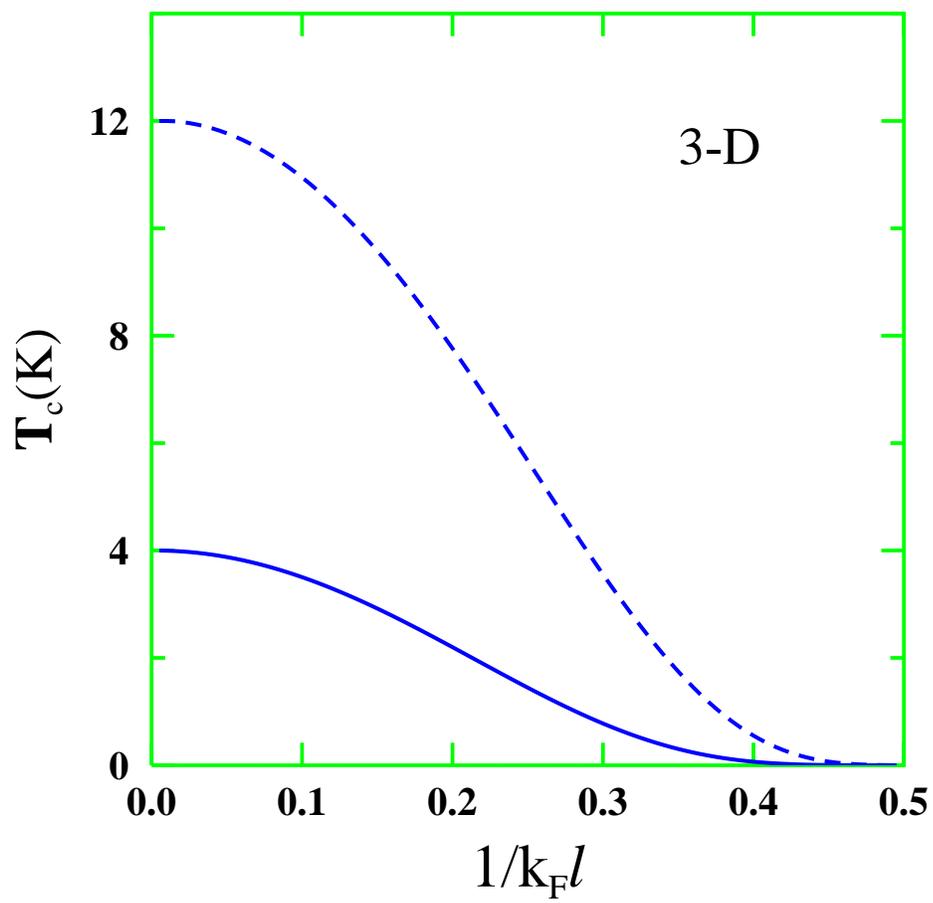
\begin{figure}
\caption{ Variation of $T_{c}$ with disorder parameter $1/k_{F}\ell$
(which represents ordinary impurity concentration) for 3-dimensional
superconductors with $T_{co}$ = 4 and 12 K. }
\end{figure}

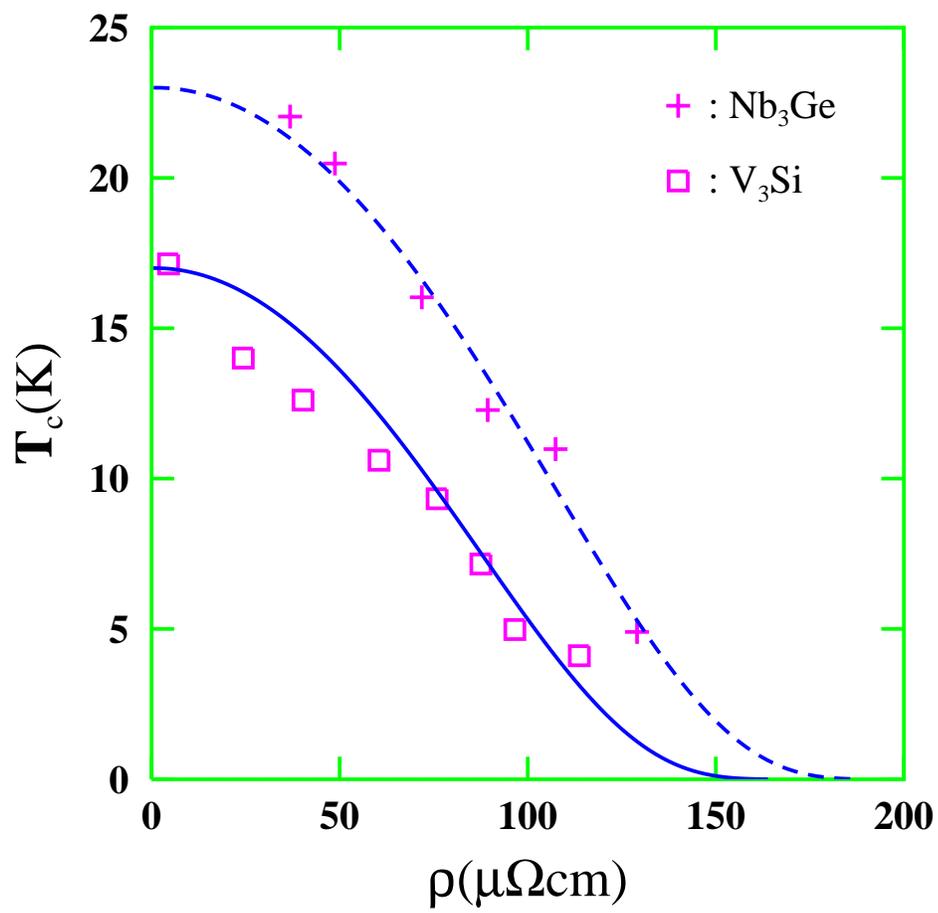
\begin{figure}
\caption{ Calculated $T_{c}$'s versus resistivity $\rho$ for 3-dimensional
Nb$_3$Ge (dotted line) and V$_3$Si (solid line). Experimental data are
from Ref. 46. }
\end{figure}

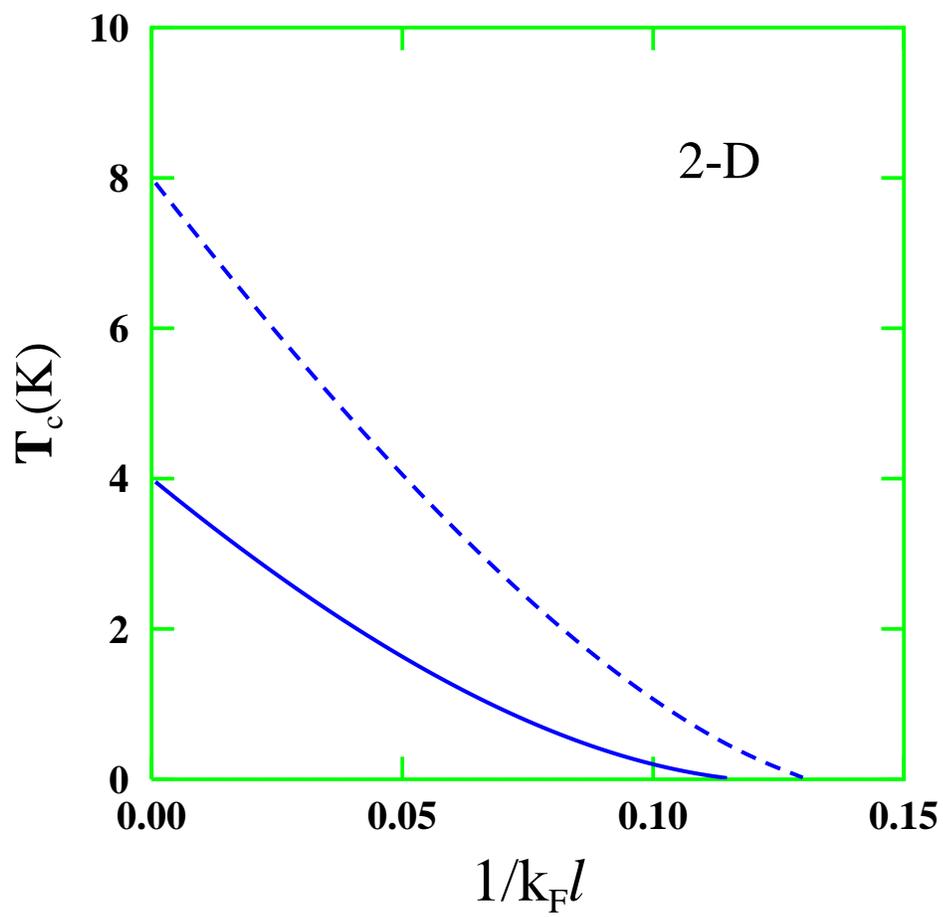
\begin{figure}
\caption{ Variation of $T_{c}$ with disorder parameter $1/k_{F}\ell$ 
for 2-dimensional superconductors with $T_{co}$ = 4 and 8 K. }
\end{figure}

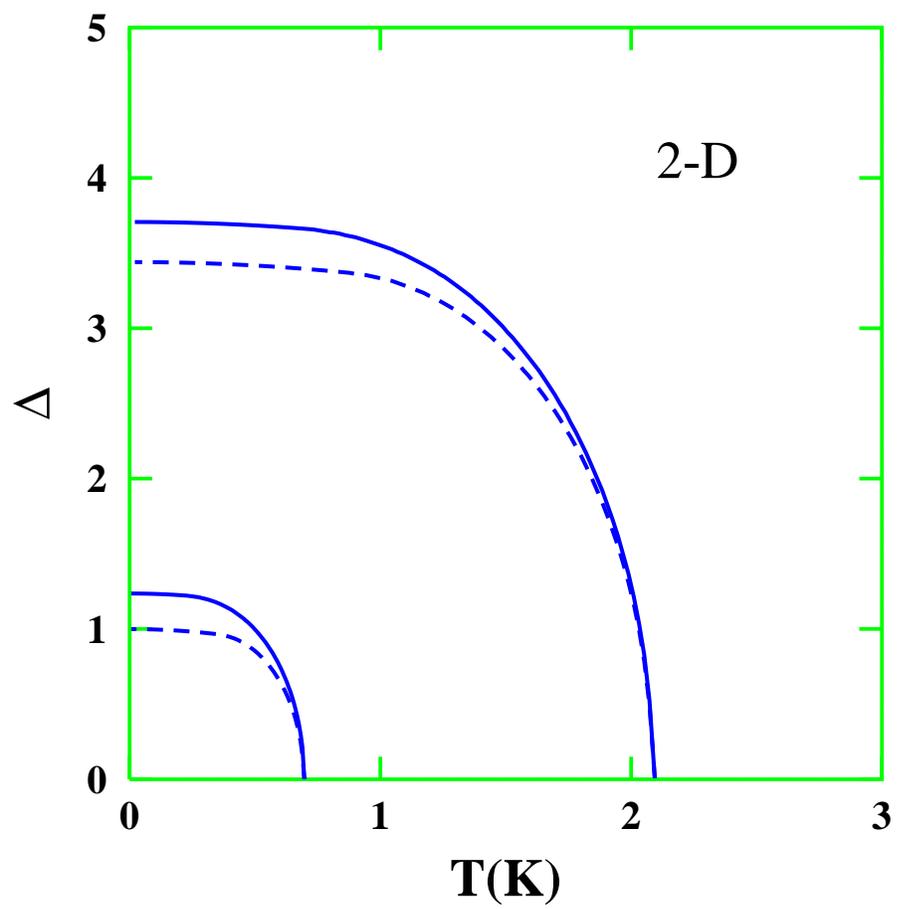
\begin{figure}
\caption{ Temperature dependence of the gap parameter for 2-dimensional
superconductors. The values of $1/k_{F}\ell$ = 0.039 and $R_{\Box}$ =
2000 $\Omega$ are chosen for the upper curves, while $1/k_{F}\ell$ =
0.078 and $R_{\Box}$ = 3000 $\Omega$ for the lower curves,
with $T_{co}$ = 4 K fixed. The $T$-independent (T-dependent) effective
electron-phonon coupling is used for the solid (dotted) lines. }
\end{figure}

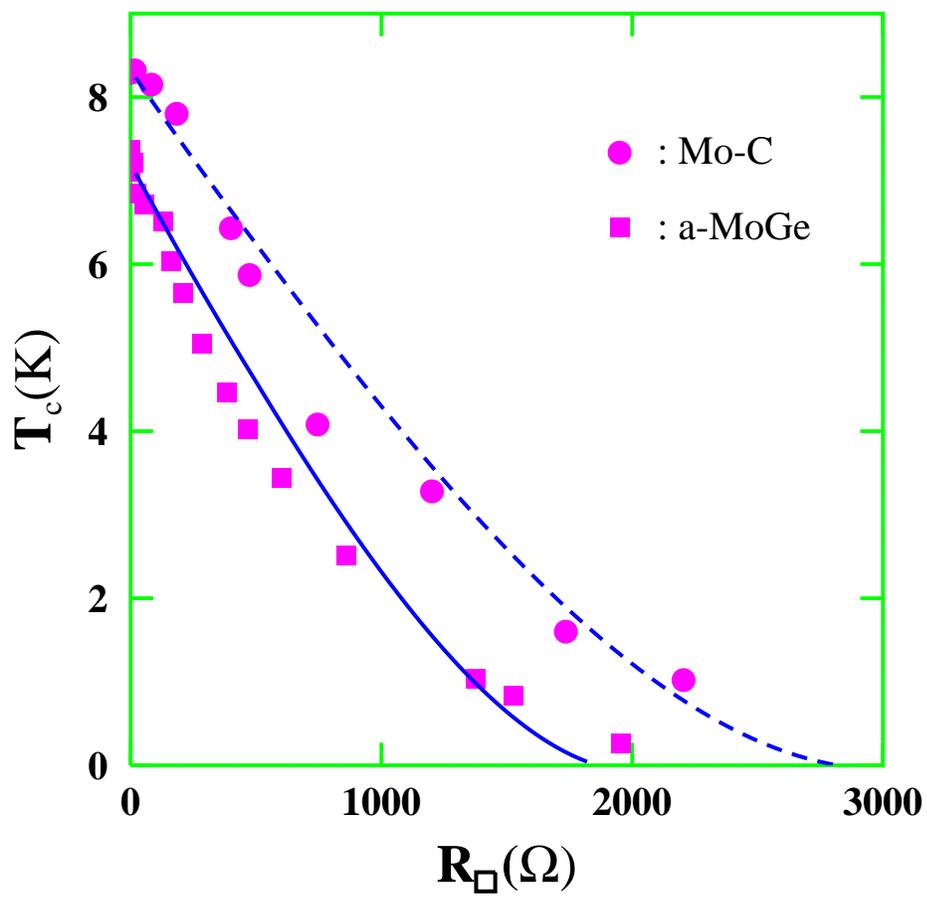
\begin{figure}
\caption{ Calculated $T_{c}$'s versus sheet resistance $R_{\Box}$
for a-MoGe (solid line) and Mo-C (dotted line) thin films. Experimental
data for a-MoGe and Mo-C are from Refs. 12 and 49, respectively. }
\end{figure}
\end{document}